\documentclass[prb,aps,twocolumn,amsmath,amssymb]{revtex4-2}

\usepackage{multirow}
\usepackage{amssymb}
\usepackage{amsmath}
\usepackage{mathrsfs}
\usepackage{graphicx}
\usepackage{amsfonts}
\usepackage{color}
\usepackage{bm}
\usepackage{xspace,soul}
\usepackage{hyperref}

\newcommand{\ket}[1]{\ensuremath{\left|#1\right\rangle}}
\newcommand{\bracket}[2]{\ensuremath{\left\langle #1 \middle| #2 \right\rangle}}

\usepackage[normalem]{ulem}

\newcommand{\blue}[1]{\textcolor{black}{#1}}

\usepackage{lineno}
\usepackage{graphicx,subfigure}

\begin{document}

\title 
{Enhancing sensitivity to rotations with quantum solitonic currents}
%{Angular momentum fractionalization and the quantum advantage of entangled solitonics currents}

\author{P. Naldesi$^1$
\footnote{Electronic address: \texttt{piero.naldesi@lpmmc.cnrs.fr}}, 
J. Polo$^{1,2}$, 
V. Dunjko$^3$, H. Perrin$^4$,
M. Olshanii$^{3}$, 
L. Amico$^{5,6,7,8,9}$, and 
 A. Minguzzi$^1$ }

\affiliation{$^1$ Univ.~Grenoble-Alpes, LPMMC, F-38000 Grenoble, France and CNRS, LPMMC, F-38000 Grenoble, France}
\affiliation{$^2$ Quantum Systems Unit, Okinawa Institute of Science and Technology Graduate University, Onna, Okinawa 904-0495, Japan}
\affiliation{$^3$ Department of Physics, University of Massachusetts Boston, Boston, MA 02125, USA}
\affiliation{$^4$ Laboratoire de physique des lasers, CNRS UMR 7538 and Universit\'{e} Paris 13 Sorbonne Paris Cit\'e, 99 av. J.-B. Cl\'ement, F-93430 Villetaneuse, France}
\affiliation{$^5$ Dipartimento di Fisica e Astronomia, Via S. Sofia 64, 95127 Catania, Italy}
\affiliation{$^6$ Centre for Quantum Technologies, National University of Singapore, 3 Science Drive 2, Singapore 117543, Singapore}
\affiliation{$^7$ MajuLab, CNRS-UNS-NUS-NTU International Joint Research Unit, UMI 3654, Singapore}
\affiliation{$^8$ CNR-MATIS-IMM \& INFN-Sezione di Catania, Via S. Sofia 64, 95127 Catania, Italy}
\affiliation{$^9$ LANEF \textit{`Chaire d'excellence'}, Universit\'e Grenoble-Alpes \& CNRS, F-38000 Grenoble, France}

\begin{abstract}
Quantum mechanics is characterized by quantum coherence and entanglement. After having discovered how these fundamental concepts govern physical reality, scientists have been devoting intense efforts to harness them to shape future science and technology\cite{dowling2003quantum}. This is a highly nontrivial task because most often quantum coherence and entanglement are difficult to access\cite{acin2018quantum}. Here, we demonstrate the enhancement of sensitivity of a quantum many body system with specific coherence and entanglement properties. 
Our physical system is made of strongly correlated attracting neutral bosons flowing in a ring-shaped potential of mesoscopic size. Because of attractive interactions, quantum analogs of bright solitons are formed \cite{kanamoto2005symmetry,calabrese2007correlation,Naldesi2018}. As a genuine quantum-many-body feature, we demonstrate that angular momentum fractionalization occurs. As a consequence, the matter-wave current in our system can react to very small changes of rotation or other artificial gauge fields. We work out a protocol to entangle such quantum solitonic currents, allowing to operate rotation sensors and gyroscopes to Heinsenberg-limited sensitivity.
\end{abstract}
\maketitle

Phase coherence is one of the most pervasive concepts in science and technology. In classical physics, coherence leads to interference. With classical interference, we fabricated devices for every-day life, for example, to manipulate sound waves or for audio-video transmissions.
With quantum mechanics, we discovered that also massive particles can be coherent. 
The technological progress that followed has had a disrupting impact in shaping the world as we know it now, with electronics, computer science, photonics, etc. 
%atom interferometry \cite{Gustavson1997,Abend2016} 
The nature of quantum coherence, though, poses challenging questions when it is referred to many-particle systems because of entanglement-induced non-local correlations. Such features have been of central importance in quantum optics~\cite{mandel1995optical}, mesoscopic physics~\cite{pichard1991quantum}, and quantum material science~\cite{anderson2018basic}, and they are now at the heart of
quantum technology. Indeed, the defining goal of quantum technology is to realize new concepts of quantum devices and simulators harnessing quantum coherence and entanglement~\cite{dowling2003quantum}.

A natural way to access the resources needed for quantum technologies is to refer to quantum many-body systems. 
Several options have been studied so far, with the different choices implying a quantum technology with different features. 
For example, superconducting circuits and circuit QED rely on the quantum coherence resulting from the specific electronic (pairing) correlations occurring in superconductors~\cite{wendin2017quantum}; with similar logic, quantum devices robust to imperfections and noise have been conceived as based on the braiding properties of the quasiparticles in the topological matter as provided e.g. by quantum Hall systems or other topological matter~\cite{nayak2008non}. 

Even though entanglement and quantum coherence are certainly present in many-body systems, it is very challenging to demonstrate how such genuine quantum resources can be of operational value in quantum technology~\cite{amico2008entanglement,Plenio.RevModPhys.89.041003}.
%In the specific yet paradigmatic problem of computing the long-time dynamics of many-body quantum systems,
In particular, it is still an open question to prove the quantum advantage of quantum simulators over classical ones~\cite{acin2018quantum,google_quant}.
Such questions are of key importance also for quantum sensing, to explore the fundamental limits of metrology~\cite{chua2014quantum,taylor2013biological,degen2017quantum,pezze2016non}.
In precision measurement, many-body correlations have recently been used in optical lattice clocks to prepare isolated atoms~\cite{Campbell2017}, allowing in turn to measure many-body effects with clock precision~\cite{Goban2018}. With atomic ensembles, massive particle entanglement has enabled a noise reduction of factor 100 in a microwave clock system \cite{Hosten2016}  

\begin{figure}[!!!h]
 \includegraphics[width=0.95 \columnwidth ]{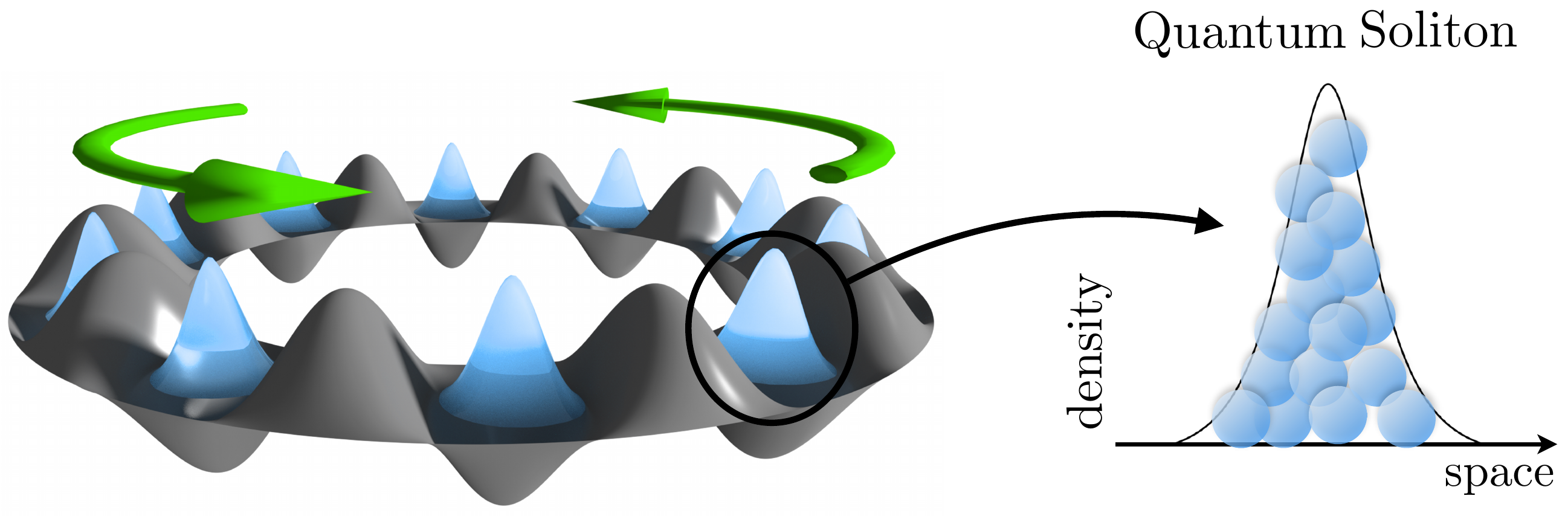}
 \caption{Schematic representation of the system. Left: Ring lattice of bosons with attractive interactions subjected to an artificial gauge field inducing matter-wave currents (arrows). Attractive interactions give rise to the formation of many-body bound states, ie quantum analogs of bright solitons, where many particles are clustered together (right).}
 \label{fig:scheme_ring_lattice}
 \end{figure}

%In this paper, we demonstrate the quantum advantage of entangled {current states of bosonic particles}, setting the basis for %quantum sensors with enhanced performances.
%At the same time, we provide a clear instance of how quantum technology can disclose new fundamental aspects of many-body %physics.
Our system is made of %entangled
attracting neutral bosonic atoms
flowing in a ring-shaped lattice potential of mesoscopic size which sustains a neutral persistent current flow (see Fig.~\ref{fig:scheme_ring_lattice}). 
As physical implementation of such a system, we propose ultra-cold atoms~\cite{bloch2008coherence}, with the new twist provided by atomtronics~\cite{Amico_NJP,Amico_Atomtronics}.
\blue{
  In contrast with continuous systems, lattice rings provide a characteristic energy-band structure, displaying bendings, foldings and energy gaps. Such features lead to a specific  protection of the bright solitons\cite{Naldesi2018}. On the other hand, we shall see that the lattice system provides a nontrivial generalization of a  theorem due to Leggett \cite{nanoelectronics1991dk}  that predicts
  %implies
  the characteristic response to an applied (artificial) magnetic field in quantum  rings.
}
%
%
%thus
%can exploit quantum technology to disclose new fundamental basic science.
%
 While our discussions apply to any type of artificial gauge fields \cite{dalibard2011colloquium}, in the following we will refer to the case of an artificial gauge field induced by a global rotation at angular frequency $\Omega$. 
For such systems, it is found that the induced angular momentum increases in quantized steps as a function of $\Omega$ \cite{moulder2012quantized,Wright2013}; correspondingly, the amplitude of persistent currents displays periodic oscillations with $\Omega$ \cite{byers1961theoretical,onsager1961magnetic}, with a periodicity that Leggett proved to be fixed by the effective flux quantum of the system, irrespective of particle-particle interactions \cite{nanoelectronics1991dk}. 

 Below, we demonstrate that for strongly correlated one-dimensional bosons with attractive interactions, the very nature of flux quantum is nontrivial, due to the formation of many-body bound states. 
This feature has dramatic effects on the persistent current 
that oscillates with a periodicity $N$ times smaller than in the standard case corresponding to repulsive interactions. 
Remarkably, the periodicity depends on interaction, which leads to an extension of the Leggett theorem.
We show how our system can be harnessed
to construct specific entangled states of persistent currents characterised by sensitivity to the effective magnetic field reaching the Heiseberg limit (quantum advantage).

Before treating the general case of the lattice ring, we will first assume that the density $N/L$ of bosons, where $N$ is the particle number and $L=2\pi R$ is the perimeter of the ring of radius $R$, is small enough to describe the system through the continuous Bose-gas integrable theory or equivalently the Lieb-Liniger model~\cite{amico2004universality}. 
For such systems, we can apply exact results~\cite{Lieb_1963}. 

In the frame rotating at frequency $\Omega$, 
the Lieb-Liniger Hamiltonian reads $\hat{\mathcal{H}}_{LL}=\sum_{j=1}^N\frac{1}{2m} p_j^2 - \Omega L_z + g \sum_{j<l} \delta(x_j-x_l)$, where $m$ and the $p_i$'s are respectively the mass and the momentum of each particle, $L_z=\sum_{j=1}^N L_{z,j}$ is the total angular momentum of the $N$ particles and $g$ is the interaction strength. The Lieb-Liniger Hamiltonian can be recast to
\begin{equation}
\hat{\mathcal{H}}_{LL}=\sum_{j=1}^N\frac{1}{2m}\bigg(p_j-m\Omega R\bigg)^2 + g \sum_{j<l} \delta(x_j-x_l) +E_\Omega ,
\label{LL}
\end{equation}
with a constant $E_\Omega=-N m \Omega^2 R^2/2$. Here, we assume periodic boundary conditions.
Using a transformation to
Jacobi coordinates (see Methods) $\xi_l$ and their canonically conjugate momenta $Q_l$, where 
$\xi_N=X_{\textrm{CM}}$ and $Q_N=P_{\textrm{CM}}$ where $X_{\textrm{CM}}=(1/N)\sum_j x_j$ and $P_{\textrm{CM}}=\sum_j p_j$ are, respectively, the coordinate and momentum of the center of mass, we find that in the Hamiltonian~(\ref{LL}) only the center-of-mass momentum is coupled to the artificial gauge field $\Omega$.
Correspondingly, the many-body wavefunction can be written as $ \Psi(x_1,...,x_N)\!=\!e^{i(P_{\textrm{CM}}-N m \Omega R) X_{\textrm{CM}}/\hbar}\chi_{\text{relative}}(\xi_{1},...,\xi_{N-1}) $. In this case, $P_{\textrm{CM}}\!=\!\ell \hbar / R $ can take any value allowed by quantization of momentum in the ring ($\ell$ being an integer).
The ground-state energy reads
$ E_{GS}\!=\! \frac{1}{2 N m } \left(P_{\textrm{CM}} - N m \Omega R \right)^2 +E_{int}$,
where $E_{int}$ is the interaction energy of the fluid, which does not depend on $\Omega$.

For repulsive interactions, independently of the interaction, $ E_{GS}$ results periodic in $\Omega$ with period $\Omega_0\!=\!\hbar/mR^2$ (see Methods). Therefore, the persistent current in the rotating frame $I_p\!=\!-(\Omega_0/\hbar)\partial E_{GS}/\partial \Omega$ reflects the center-of-mass quantization, and displays the characteristic sawtooth behaviour versus $\Omega$ \cite{nanoelectronics1991dk}, corresponding to a staircase behaviour of angular momentum $L_z$.

For attractive interactions the ground state is a many-body bound state, {\it i.e.} a 'molecule' made of $N$ bosons, corresponding to the quantum analog of a bright soliton \cite{kanamoto2005symmetry,calabrese2007correlation,Naldesi2018}. \blue{This picture arises from the exact Bethe ansatz solution; within the regime of validity of the string hypothesis\cite{calabrese2007correlation} (see Methods)} the ground state energy for arbitrary $\Omega$ reads
\begin{equation}
 E_{GS}= \frac{\hbar^2}{2 M R^2}\left(\ell -N\frac{\Omega}{\Omega_0}\right)^2 - {{N (N^2-1)g^2}\over{12}},
 \label{egs-ll}
\end{equation}
where the second term accounts for the interaction energy $E_{int}$. The above equation shows that attracting bosons behave as a single massive object of mass $M=Nm$ under the effect of the artificial gauge field. The energy displays a $1/N$-periodicity as a function of the artificial gauge field, $\Omega$, in units of $\Omega_0$ corresponding to {\it fractionalisation} of angular momentum per particle. In analogy to the fractional quantum Hall effect, in our system, the elementary particles carrying a fraction of quantum of angular momentum are parts of composite objects. We shall see, however, that our composite object displays a very specific dependence on the interplay between interaction and system size. 
%\red{JOAN: recent paper I have found https://arxiv.org/pdf/2004.13504.pdf}

 {To this end, we discuss the general non-integrable case in which the lattice effects are relevant. We assume} that the bosons dynamics is entailed by the Bose-Hubbard Model (BHM):
\begin{equation}
\hat{\mathcal{H}}_{BH}= \sum_{j=1}^{N_s} \frac{U}{2} n_{j}\left( {n}_{j}-1\right) -J\left( e^{-i \tilde{\Omega}}a_{j}^{\dagger }a_{j+1}+\text{h.c.} 
\right) ,
\label{BHH}
\end{equation}
where $a_{j}$ and $a_{j}^{\dagger }$ are site $j$ annihilation and creation Bose operators and $n_{j}\!=\!a_{j}^{\dagger }a_{j}$. The parameters $J$, $U<0$ in \eqref{BHH} are respectively the hopping amplitude and the strength of the on-site interaction, $N_s$ being the number of sites in the ring lattice and $\tilde{\Omega}\doteq 2\pi \Omega/(\Omega_0N_s)$ for brevity.

%~ \onecolumngrid
%~ \begin{center}
 %~ \begin{figure}[tbh!]
 %~ \includegraphics[width=0.85 \columnwidth ]{fig_2.pdf}
 \begin{figure*}[tbh!]
 \includegraphics[width=1.7 \columnwidth ]{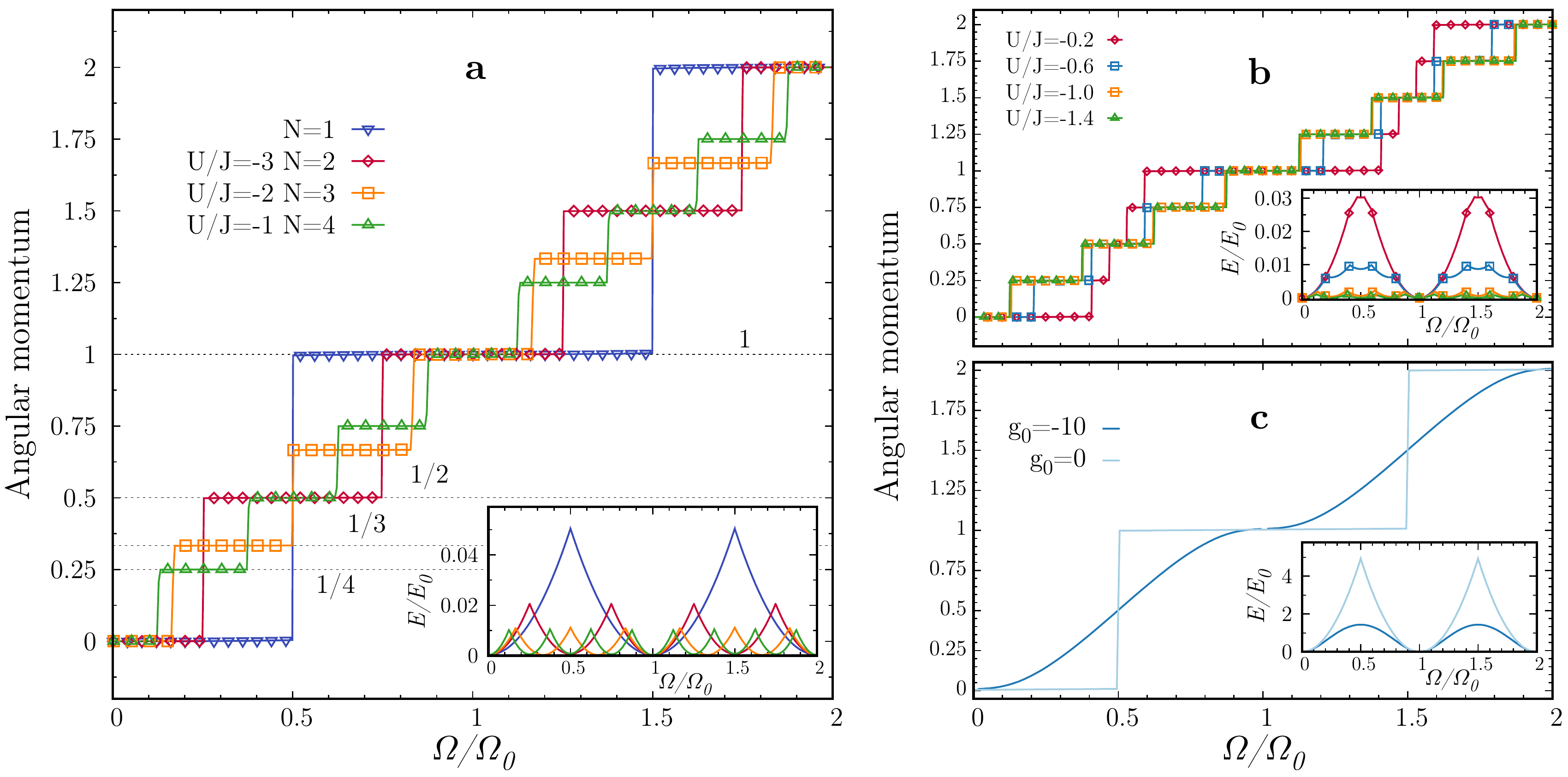}
\caption{
\textbf{Fractionalisation of angular momentum in a Bose gas with attractive interactions.}
Average angular momentum per particle (main) and ground-state energy (inset) for bosons on a lattice ring as a function of artificial gauge field, from numerical exact diagonalization calculations: \textbf{a}) at varying particle number, for chosen values of interaction strength as indicated on the figure, \textbf{b}) for various values of interaction strength at fixed $N=4$. Panel \textbf{c}) shows the corresponding predictions from the mean-field Gross-Pitaevskii equation for zero and finite attractive interactions indicated by the dimensionless parameter $g_0=m g L/\hbar^2 $.
The angular momentum per particle is obtained as $\displaystyle{\hbar \ell = {{M_{eff} }\over{M}} \left ({{ \hbar I_p}\over{\Omega_0}} + \Omega \left.{{\partial^2 E_{GS}}\over{\partial \Omega^2}}\right|_{\Omega=0} \right ) }$, with $M_{eff}$ being the effective mass of the bound state in the lattice.}
 \label{fig:current_gaguefield_interactions_particles}
 \end{figure*}
 %~ \end{figure}
%~ \end{center} 
%~ \twocolumngrid

We point out that, for the lattice model~(\ref{BHH}), the center-of-mass and relative coordinates do not decouple (for any finite interaction).
As an effect, the internal structure of the many-body bound state is affected by the interplay between interaction and artificial gauge field $\Omega$ (since $P_{CM}$ depends on $\Omega$, and the internal structure depends on $P_{CM}$).
\blue{
  Here, we find that the periodicity of the persistent current for lattice rings
  %structures REPETITION WITH 'structure above'
  does depend on interaction.
We remark that such a 'non-perfect' fractionalization (see Fig.~\ref{fig:current_gaguefield_interactions_particles}(b)) is observed for  solitons that are  properly formed in the system (i.e. when the system size is larger than the correlation length of the density-density correlations). % we conclude that this is not a  finite size effect.
}%
%
%This feature has a profound influence on the persistent current: {\it in contrast with the continuous theory, {the periodicity of the persistent current for lattice ring attracting bosons does depend on interactions}}.
Fig.~\ref{fig:current_gaguefield_interactions_particles} shows our numerical results (confirmed by exactly solving the BHM in the 2-particle sector--see Methods) for the ground-state energy, persistent currents and angular momentum: also in the lattice nonintegrable case the $1/N$ periodicity in $\Omega/\Omega_0$ of the persistent currents emerges, as well as fractionalization of angular momentum.
Indeed, these features, though, are affected by the interplay between system size and interaction strength.
The $1/N$ periodicity is found when interactions are sufficiently large: In these conditions, the 'size of the many-body bound state', defined as the typical decay length of the density-density correlations \cite{Naldesi2018}, is smaller than the size of the system.
Upon decreasing the interactions, the many-body bound state spreads more and more over the sites {making the solitonic nature of the state less and less pronounced (see Methods).
We remark that all the observed features are {\em purely quantum many-body effects tracing back to specific quantum correlations}:
Indeed, mean-field Gross-Pitaevskii equation (corresponding to a non-entangled ground state) provides persistent currents displaying no fractionalization, independently on the strength of the interaction (see Fig.~\ref{fig:current_gaguefield_interactions_particles}, c).

Remarkably, the afore discussed angular momentum fractionalization and persistent current periodicity emerge in the time-of-flight (TOF) distributions of the atoms after releasing the trap confinement {and switching off interactions. We obtain it from $n(\mathbf{k}) = |w(\mathbf k)|^2 \sum_{j,l} e^{i \mathbf{k}\cdot(\mathbf{x}_j-\mathbf{x}_l)} \langle a^\dagger_j a_l\rangle$,
 where $\mathbf{x}_j$ indicate the position of the lattice sites in the plane of the ring and $w(\mathbf k)$ is the Fourier transform of the Wannier function of the lattice~\cite{amico2005quantum}.}  
Instead of the characteristic wide $\ell$-dependent minimum ('hole') arising for zero or repulsive interactions \cite{moulder2012quantized,Wright2013},
 we find no clear hole at $\mathbf k=0$ for the attractive case --Fig.~\ref{fig-TOF}. Such a feature is due to the reduction of coherence implied by the solitonic many-body bound state.
Despite the seemingly featureless momentum distribution, we find that {\it fractional steps of the mean-square radius of the distribution for $\Omega/\Omega_0=\ell/N$} \cite{Brentin2004}. {\it This effect provides the univocal signature of $1/N$ fractionalization of angular momentum in the presence of a many-body bound state. }

 \begin{center}
 \begin{figure}[t]
 \includegraphics[width=0.85 \columnwidth ]{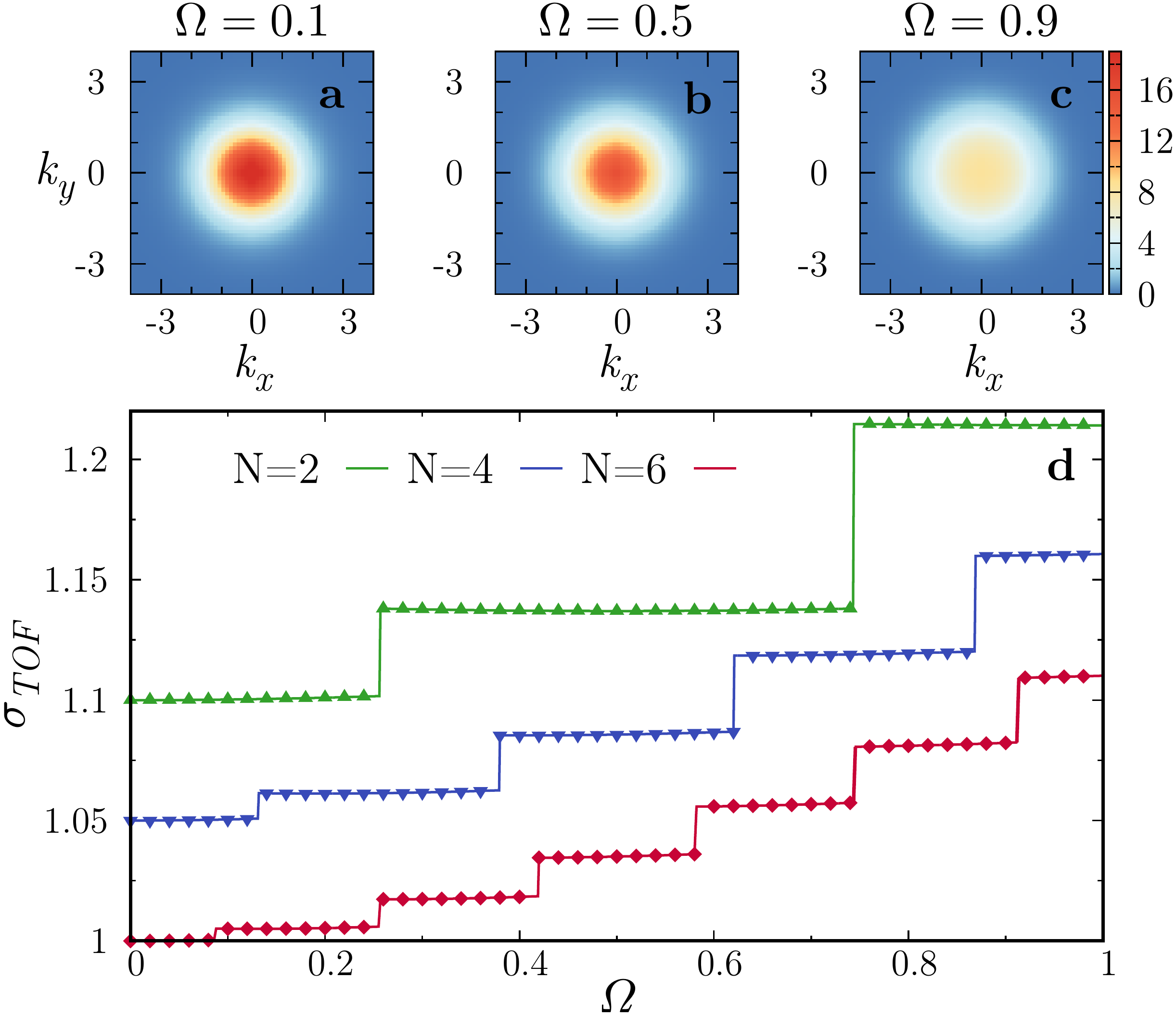}
 \caption{\textbf{Time of flight expansion of the Bose gas after releasing the atoms from the ring trap.}
 \textbf{a-b-c} Density plot of the TOF expansion for different values of the artificial gauge field
 in a system of $N=4$ particles and $L=11$ lattice sites. \textbf{d} Renormalized width $\sigma_{\textrm{TOF}}/\sigma_{\textrm{TOF}}(\Omega=0)$ of the time of flight density distribution, $n(\textbf{k})$, for different number of particles, and interactions $U/J=-0.6$, $U/J=-1$ and $U/J=-3$ respectively. For the sake of graphical clarity, each curve is offset by $0.05$. Note how the TOF density distribution width abruptly changes with the increase of the strength of the artificial gauge field, and how the sensitivity proportionally increases with the number of particles. In all the calculations we have approximated the Wannier functions with Gaussians functions with width $\delta=a/\sqrt{2 \pi}$ with $a$ the lattice spacing.}
 \label{fig-TOF}
 \end{figure}
\end{center}

We finally demonstrate how the scenario above can be harnessed to construct entangled states of different current states with quantum advantage for atom interferometry. Indeed, our Hamiltonians Eqs.(\ref{LL}), (\ref{BHH}) commute with the total angular momentum. Therefore, to entangle states with different angular momentum, the rotational invariance of the system needs to be broken. In the following, we propose a specific protocol leading to the creation of such a state: The ring is interrupted by a localized barrier of strenght $\Delta_0$, and the artificial gauge field is quenched from $\Omega=0$ to $\Omega=\Omega_0/2$. Remarkably, this procedure {\em dynamically entangles} the angular momentum state at $\Omega=0$, ie $L_z=0$, with the one at $\Omega=\Omega_0$, ie $L_z=N$ (see again Fig.~\ref{fig:current_gaguefield_interactions_particles}), yielding  $\ket{\psi}_{NOON} = \frac{1}{\sqrt{2}} \left( \ket{L_z=0} + \ket{L_z=N} \right)$ when the current reaches the half of its maximum value. 
\blue{
We note that such entangled states are superposition of current states, which are dual to the ``NOON'' states defined in the  particle-number Fock basis \cite{Polo_2018}. 
}
The response of such a state to an external rotation is $\ket{\psi(\phi)} =e^{i \phi \hat{L}_z/\hbar} \ket{\psi}_{NOON}$, and the quantum Fisher information \cite{Caves1994,Pezze2009} $F_\mathcal{Q} = 4 \left(\bracket{\psi'(\phi)}{\psi'(\phi)} - |\bracket{\psi'(\phi)}{\psi(\phi)}|^2 \right)$, being $\ket{\psi'(\phi)}= \partial \ket{\psi(\phi)}/\partial \phi$. For our state we find $F_\mathcal{Q}\sim N^2$, ie it reaches the Heisenberg limit - see Fig.~\ref{fig:fisher}. The corresponding sensitivity $\delta \phi$, therefore, is 
\begin{equation}
\delta \phi \ge \frac{1}{(F_\mathcal{Q})^{1/2}} =\frac{1}{ N} ,
\end{equation}
This shows that 
{\it entangled states of quantum solitons with different angular momenta lead
 to a quantum advantage of the sensitivity.}
 }

Summarizing, we have demonstrated that attracting bosons on a ring display fractionalization of angular momentum.
On the fundamental level, such feature represents a remarkable extension of well known predictions due to Byers-Yang-Onsager-Leggett \cite{nanoelectronics1991dk,byers1961theoretical,onsager1961magnetic}: The many-body bound-state nature of the ground state of attractive bosons implies fractional angular momenta per particle; interactions do not change the fractionalization on a continuous ring but they do affect it in the generic (lattice) system in which also the relative coordinate of the particles are sensitive to $\Omega$. 
Such features are due to the entanglement in the ground state: the effect vanishes in the Gross-Pitaevskii limit in which the many-body wave function describes a factorized state.
The $1/N$ fractionalization can be observed experimentally by studying the system's momentum distribution; the observation of such effect would provide the evidence of the formation of many-body quantum solitons beyond the Gross-Pitaevskii mean-field regime.

\begin{figure}[h!]
 \includegraphics[width=\columnwidth ]{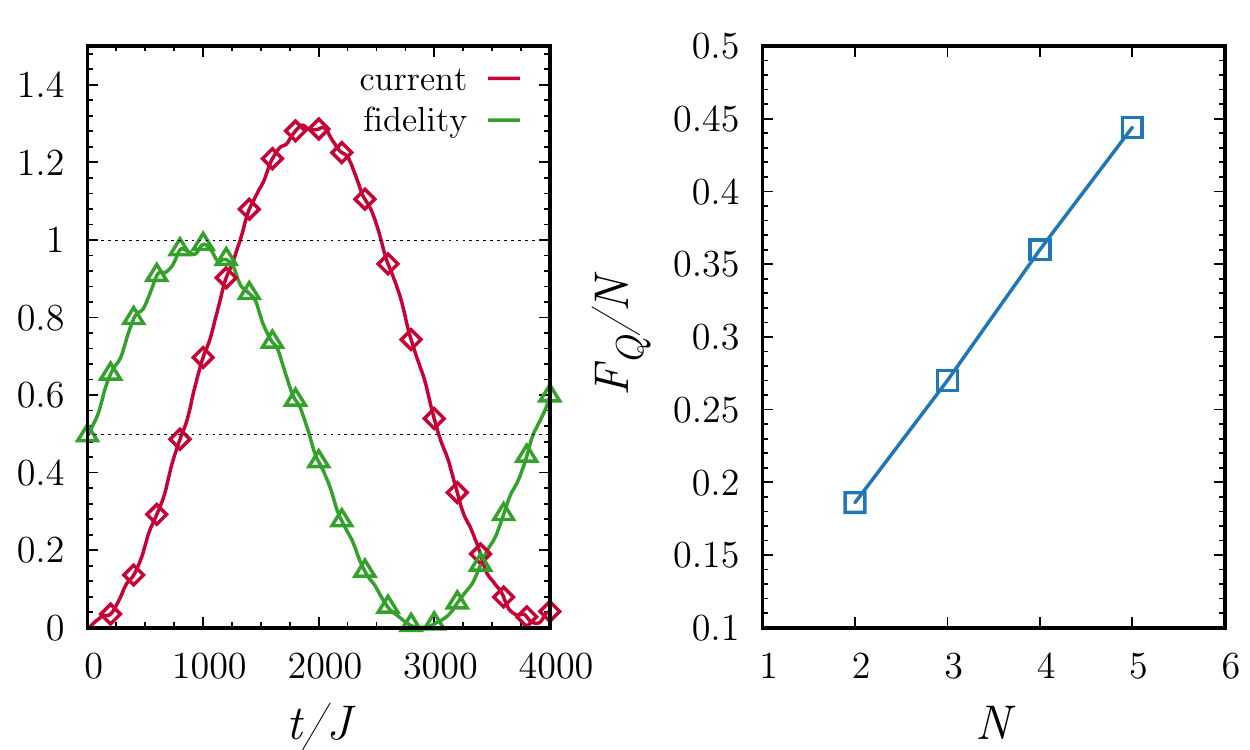}
 \caption{\textbf{Creation of entangled states of angular momentum with quantum solitons}. \textbf{a} Exact many-body dynamics of the current (in units of the hopping constant $J$) following a quench from $\Omega/\Omega_0\!=\!0$ to $\Omega/\Omega_0\!=\!1/2$.
Here we set $L=28$, $N=3$, $U/J=-0.51$ and $\Delta_0/J=0.015$. At one quarter of oscillation period, the superposition $\ket{\psi}\!=\!\frac{1}{\sqrt{2}} \left( \ket{L_z\!=\!0} + \ket{L_z\!=\!N} \right)$ is formed, with a fidelity very close to 1. No fine tuning of parameter is required. \textbf{b} Scaling of the quantum Fisher information with particle numbers, showing that it reaches the Heinsenberg limit $F_Q\propto N^2$. The system parameters are described in the supplementary material.}
 \label{fig:fisher}
\end{figure} 

We note that, because of the formation of quantum solitons, an enhanced control on $N$ in the experiments is expected; in the lattice such value is protected by a finite gap\cite{Naldesi2018}. The fractionalization of the angular momentum can define protocols to measure the number of particles in cold atoms experiments. 
Our results yield a $N$-factor enhancement in the sensitivity of attracting bosons to an external field. We have provided a protocol to prepare a superposition of current states, explicitely exploiting the strong correlations, and we demonstrated that this state has a quantum Fisher information scaling as $N^2$, thus allowing to reach the Heisenberg limit in atomic interferometry.\\

{\bf Acknowledgments}. The Grenoble LANEF framework (ANR-10-LABX-51-01) is acknowledged for its support with mutualized infrastructure. We thank National Research Foundation Singapore and the Ministry of Education Singapore Academic Research Fund Tier 2 (Grant No. MOE2015-T2-1-101) and ANR SuperRing (ANR-15-CE30-0012) for support. LPL is a member DIM SIRTEQ (Science et Ing\'enierie en R\'egion \^Ile-de-France pour les Technologies Quantiques).\\

{\bf Author contributions}. PN and JPG carried out most of calculations. VD, MO and HP analysed the device as rotation sensor and the relevant experimental observables. LA and AM conceived and supervised the work and drafted the manuscript. All the authors discussed the results and iterated the manuscript.\\

{\bf Materials and Correspondence}. Correspondence and requests for materials should be addressed to PN (piero.naldesi@lpmmc.cnrs.fr).

\bibliographystyle{apsrev4-2}

\vspace{1cm}
{\bf Methods}
\appendix

\section{Separation of center-of-mass and relative coordinates}

We detail here the coordinate transformation to center-of-mass and relative coordinates. We introduce the Jacobi coordinates
\begin{align*}
\left(\begin{array}{c} y_{1} \\ y_{2} \\ \dots \\ y_{N} \end{array}\right) = M_{\text{Jac}} \cdot \left(\begin{array}{c} x_{1} \\ x_{2} \\ \dots \\ x_{N} \end{array}\right)
\,\,,
\end{align*}
with the Jacobi matrix given by
\begin{align}
M_{\text{Jac}} =
 \left(
 \begin{array}{ccccccc} 
1       & -1     & 0       & 0             & \cdots     & 0
\\
\frac{1}{2}     & \frac{1}{2}   & -1      & 0             & \cdots     & 0
\\
\frac{1}{3}     & \frac{1}{3}   & \frac{1}{3}  & -1         & \cdots     & 0
\\
\cdots    &\cdots    &\cdots    &\cdots        &\cdots    & 0
\\
\frac{1}{N-1}  &\frac{1}{N-1}  &\frac{1}{N-1}  &\frac{1}{N-1}    &\cdots    &-1   
\\
\frac{1}{N}    &\frac{1}{N}  &\frac{1}{N}  &\frac{1}{N}    &\cdots    &\frac{1}{N}     
\end{array}
\right) 
\,\,.
\label{Jacobi_matrix}
\end{align}
Here, $y_{N} = X_{\text{CM}} \equiv \frac{\sum_{l=1}^{N} x_{l}}{N}$ is the center-of-mass coordinate we want to separate out.
The Jacobi matrix~(\ref{Jacobi_matrix}) is however not orthogonal (ie it is not a rotation). Nonetheless,
the matrix $M_{\text{Jac}}$ can be easily converted to a pure rotation $R_{\text{Jac}}$ via the rescaling:
\begin{align}
R_{\text{Jac}} = \mbox{diag}\left( \sqrt{\frac{1}{2}},\, \sqrt{\frac{2}{3}},\, \ldots,\, \sqrt{\frac{N-1}{N}},\, \sqrt{N} \right) \cdot M_{\text{Jac}}
\label{quasi-Jacobi_rotation}
\,\,.
\end{align}
where $\mbox{diag}(\dots)$ is a diagonal matrix, with the numbers in the parenthesis specifying the diagonal matrix elements. 
Indeed one can straightforwardly verify that $R_{\text{Jac}} \cdot R_{\text{Jac}} ^{\top} = 1$, where $R_{\text{Jac}} ^{\top}$ is a transpose of $R_{\text{Jac}} $. 

We define then the coordinates
\begin{align*}
\left(\begin{array}{c} z_{1} \\ z_{2} \\ \dots \\ z_{N} \end{array}\right) = R_{\text{Jac}} \cdot \left(\begin{array}{c} x_{1} \\ x_{2} \\ \dots \\ x_{N} \end{array}\right)
\,\,.
\end{align*}
Note that $z_{N} =\sqrt{N} X_{\text{CM}}$.

Let us now introduce one final transformation, which brings us back to $X_{\text{CM}}$ as one of the variables, while keeping the Jacobian determinant of the transformation equal to one:
\begin{align*}
&
\xi_{l} = N^{\frac{1}{2(N-1)}} z_{l}, \quad l=1,\,2,\,\ldots,\,N-1 
\\
&
\xi_{N} = \frac{1}{\sqrt{N}}z_N = X_{\text{CM}}
\,\,.
\end{align*}
This defines the relative and center-of-mass coordinates used in the main text.

By a similar procedure one can identify the transformation to the Jacobi momenta $Q_l$, canonically conjugate to $\xi_l$, where $Q_N=\sum_{j=1}^N p_j=P_{\textrm{CM}}$ is the center-of-mass momentum.
In particular, by introducing a set of momenta $\vec P_z= R_{\text{Jac}} \vec p$, with the same Jacobi matrix $R_{\text{Jac}}$ as the one used for spatial coordinates, one can show that $Q_l= \alpha P_{z_l}$ for $l=1,...N-1$ with $\alpha=N^{-1/[2(N-1)]}$, and $Q_N=\sqrt{N} P_{z_N}=P_{\textrm{CM}}$

The final Hamiltonian then reads
\begin{eqnarray}
 H&=&\sum_{j=1}^{N-1}\frac{1}{2\mu_N} Q_j^2 + V_{int}(\xi_1,...,\xi_{N-1})\nonumber \\ &+& \frac{1}{2M} \left(P_{\textrm{CM}} - N m \Omega R\right)^2
\end{eqnarray}
where $\mu_{N} \equiv N^{-\frac{1}{(N-1)}} m$ is the mass of the relative problem, $M=N m$ is the total mass.

\section{Exact Bethe Ansatz results for the continuous ring}
We start from the Lieb-Liniger model Eq.(1) of the main text, where we drop the constant $E_\Omega$:
\begin{equation}
H_{LL}=\sum_{j=1}^N\frac{1}{2m}\bigg(p_j-m \Omega R \bigg)^2 + g \sum_{j<l} \delta(x_j-x_l),
\label{LLeq}
\end{equation}

For the Lieb-Liniger model, the total momentum and energy are $P_{\textrm{CM}}=\hbar \sum_{j=1}^{N} k_j$
and $E=(\hbar^2/2m)\sum_{j=1}^{N} k_j^2$ respectively,
where the $k_j$ are obtained by solving the Bethe equations 
\begin{equation}
k_j= {{2 I_j \pi}\over{L}} + 2 \pi {{\Omega}\over{\Omega_0 L}} -\sum_\ell \arctan\left ({{k_j-k_\ell}\over{c}}\right )
\end{equation}
where $c= 2m g/\hbar^2$, $L=2 \pi R$ is the ring circumference and $I_j$ is a set of integer (semi-integer) numbers defining the state of the system. For repulsive interactions, all the $k_j$'s are real. For $2l \pi/L \le \Omega\le 2(l+1) \pi/L $, the ground states can be obtained by $I_j=-(N-1)/2+j+\ell$, with integer $\ell$, yielding a center of mass momentum given by $P_{CM}=\hbar \sum_jk_j =\ell N \hbar/R$, as readily follows by noticing that arctan$[(k_j-k_\ell)/c]$ is an odd function. 

For repulsive interactions,
the allowed values for the center of mass are integer multiples of $2 p_F$, where $p_F= \hbar N/2R$, yielding
$E_{GS}= \frac{N \hbar^2}{2mR^2} \left(\ell - \Omega/\Omega_0\right)^2 +E_{int}$
with $\Omega_0=\hbar/mR^2$.
The ground state energy hence results periodic in $\Omega$ with period $\Omega_0$ and the persistent current, obtained as $I_p=-(\Omega_0/\hbar)\partial E_{GS}/\partial \Omega$, clearly reflects the center-of-mass quantization for any value of interaction strengths.

For attractive interactions the Bethe equations of (~\ref{LLeq}) admits complex solutions and the ground state corresponds to a many-body bound state: $k_j=\kappa- i (n+1-2j) g/2$, $j=1\dots n$. Such $n$ string solutions holds also for $\Omega\neq 0$, since the scattering matrix is not affected by $\Omega$. The ground state of (~\ref{LLeq}) is made of a single $n=N$-string, yielding Eq.(\ref{egs-ll}) of the main text. 
\blue{Here we point out that string hypothesis holds for $cL\rightarrow\infty$. The finite size corrections to the string solutions (for recent references, see \cite{sakmann2005exact,sykes2007excitation}) can affect the interaction energy $E_{int}$. 
}

\section{Numerical Methods}

Here we present the numerical techniques that have been used to obtain the results presented in this paper. We solve the eigenvalue problem by writing the Hamiltonian, $\mathcal{\hat{H}}$, as a matrix $H_{ij}$ in the Fock basis. This basis is then hashed in a more efficient form \cite{Cameron_1994} in order to write the Hamiltonian in a sparse way. In particular, our numerical code is written in Python and the sparse Hamiltonian is diagonalized using ARPACK within the SciPy library. We have performed simulations with $N_s=11$ to $N_s=24$ sites and $N=2$ to $N=6$ particles, with a Hilbert space dimension up to $10^6$, for different values of the flux $\Omega/\Omega_0$. Simulations have also been benchmarked with DMRG \cite{whitedmrg} data. After solving the eigenvalue problem, the correlation function $C_{lk}=\langle a_l^{\dagger}a_k \rangle$ is calculated using the ground state of the system and is used to obtain the time-of-flight results of Fig.~3.

%%%%%%%%%%%%%%%%%%%%%%%%%%%%%%%%%%%%%%%%%%%%%%%%%%%%%%%%%%%%%%%%%%%%%%%%%%%%%%%%%%%%%%%%%%%%%%%%%%%%%%%%%%%%%%%%%%%%%%%%
%%%%%%%%%%%%%%%%%%%%%%%%%%%%%%%%%%%%%%%%%%%%%%%%%%%%%%%%%%%%%%%%%%%%%%%%%%%%%%%%%%%%%%%%%%%%%%%%%%%%%%%%%%%%%%%%%%%%%%%%
\section{Two-particle exact solution}

In the $N=2$ sector, the Bose-Hubbard model the many-body wavefunction can be obtained using the coordinated Bethe Ansatz approach.
Therefore, the ground-state energy and correlation functions can be accessed exactly.
We generalize Ref.\cite{Boschi_2014} to include the presence of an artificial gauge field in the Hamiltonian.
Here, we gauge away the Peirerls factors in the Hamiltonian and we impose twisted boundary conditions: $\hat{a}_{N_s+1}=e^{2\pi i\Omega/\Omega_0}\hat{a}_{1}$. 
A general two particle state can be written as:
 \begin{equation}
 |\phi\rangle =\sum_{j,k=1}^{N_s}\phi_{jk}\hat{a}_j^\dagger \hat{a}_k^\dagger |0\rangle
 \end{equation}
where $\phi_{jk}$ is the two-partcile wavefunction, symmetric under the exchange of $j$ and $k$, and normalized to unity. The energy of the system is found by solving the time-independent Schr\"odinger equation $\hat{H}|\phi\rangle=E |\phi\rangle$ using the Bethe Ansatz technique. In the center-of-mass and relative discrete dimensionless coordinates $X\!=\!(j+k)/2$, $x\!=\!j-k$ and $P\!=\!p_1+p_2$, $p\!=\!(p_1-p_2)/2$ the wavefunction $\phi_{jk}$ reads:
\begin{equation}\label{BAXx}
\phi_{jk}=e^{iPX}\left(a_{12}e^{ip\left|x\right|}+a_{21}e^{-ip\left|x\right|}\right).
\end{equation}
The energy eigenvalues of the two-particle system are given by ${E\!=\!-4J\cos(\frac{P+\Omega}{2})\cos (p)}$. The center of mass momentum is obtained by imposing twisted boundary conditions and quantization of the ring:
\begin{eqnarray}
P_n = \frac{2 \pi }{N_s} (n-2\Omega/\Omega_0),
\end{eqnarray}
%
%*
For the BHM the relative momentum $p$ is obtained by the condition:
\begin{equation}
 (-1)^ne^{ip(N_s+1)}=y\left(P_n,p\right)\
 \label{rap_BH1}
\end{equation}
with 
\begin{eqnarray}
y\left(P_n,p\right)\equiv\frac{
a_{21}
}
{a_{12}}
&=&
-
\frac{
\frac{U}{4 J_0}
-i\cos\left(\frac{P}{2}\right) \sin(p)
}{
\frac{U}{4 J_0}
+i\cos\left(\frac{P}{2}\right) \sin(p) 
} \; .
 \label{rap_BH2}
\end{eqnarray}

It is interesting to compare the BH and the Lieb-Liniger pictures. In the latter case, the equations to solve are
\begin{equation}
e^{ipL}=Y\left(p\right)
 \label{rap_LL1}
\end{equation}
with 
\begin{eqnarray}
Y\left(p\right)\equiv\frac{
a_{21}
}
{a_{12}}
&=&
-
\frac{
c-i p
}{
c+ip
} \; .
 \label{rap_LL2}
\end{eqnarray}
%

%%%%%%%%%%%%%%%%%%%%%%%%%%%%%%%%%%%%%%%%%%%%%%%%%%%%%%%%%%%%%%%%%%%%%%%%
%%%%%%%%%%%%%%%%%%%%%%%%%%%%%%%%%%%%%%%%%%%%%%%%%%%%%%%%%%%%%%%%%%%%%%%%

Note that, in contrast with the BH case, Eqs.(\ref{rap_LL1}), (\ref{rap_LL2}) are decoupled, ie the center of mass momentum $P$ decouples to the relative momentum. 
As a result, the imaginary part of the momentum $p$ is independent on $\Omega$; this feature implies that the periodicity of the ground state energy does not change with the interaction strength.
For the BHM, instead, $P$ and $p$ are coupled; this feature has a clear effect in the periodicty of the ground state energy. 
In conclusion, in the BHM the dependence of the periodicity on interactions is an effect of the coupling between center of mass and relative momentum.

Note that by solving Eqs.(\ref{rap_BH1}), (\ref{rap_BH2}) becomes fully determined. Thus, the time of flight images can be then readily evaluated by:
\begin{eqnarray}
n(\mathbf{k}) &=& 
\sum_{j,l=1}^{N_s} e^{i \mathbf{k}\cdot(\mathbf{x}_j-\mathbf{x}_l)} \langle a_j^\dagger a_l \rangle \nonumber \\
&=&
\sum_{j,l=1}^{N_s} e^{i \mathbf{k}\cdot (\mathbf{x}_j-\mathbf{x}_l) + i \Omega (j-l)/\Omega_0} \sum_{n}\phi^{*}_{jn}\phi_{nl}
\end{eqnarray}

%\onecolumngrid
\begin{center}
 \begin{figure}[htb]
 \includegraphics[width=1.0 \columnwidth ]{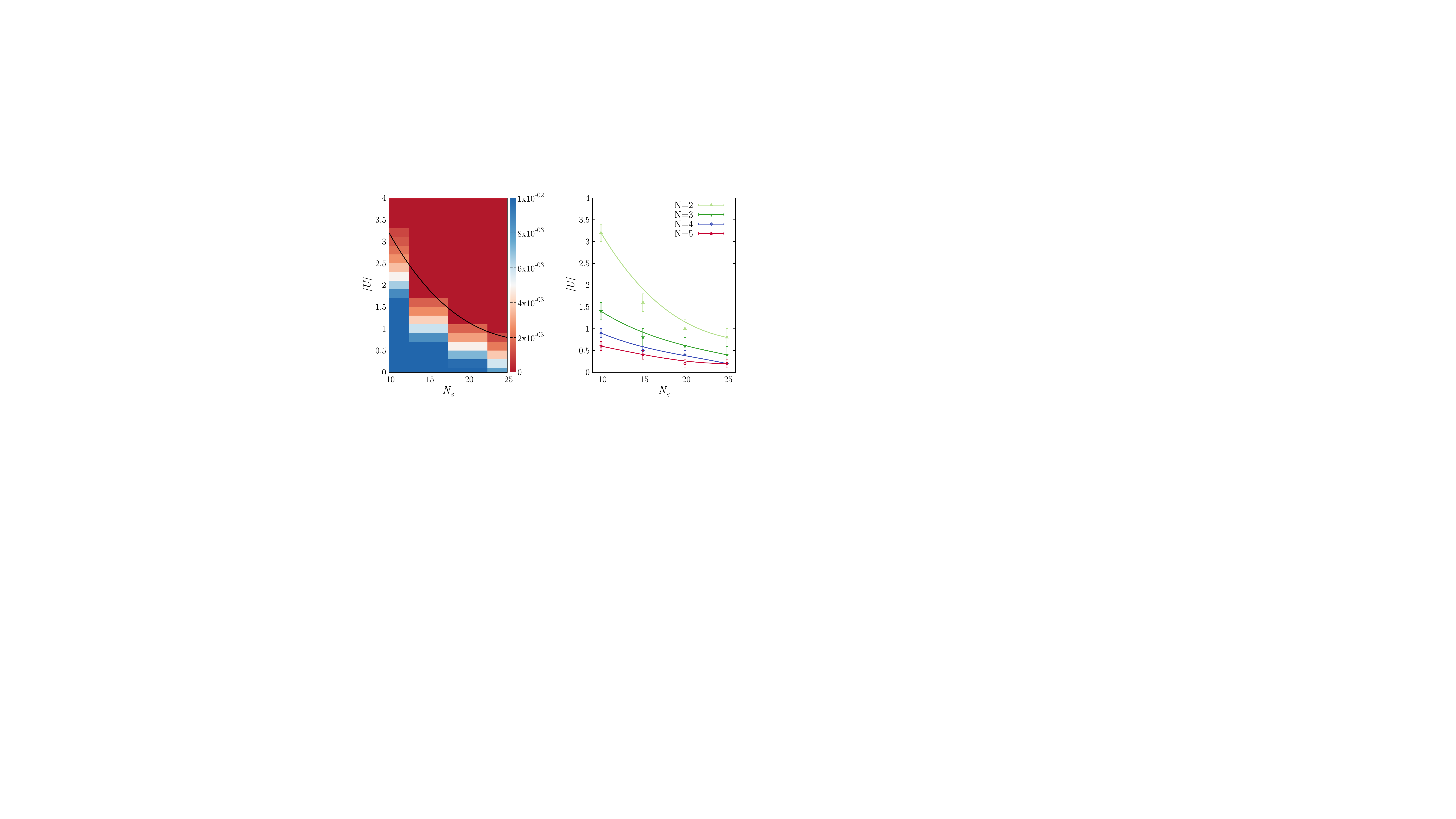}
 \caption{\textbf{a} Density plot of the renormalized energy difference between the $N$-times periodicity and the nonrotating system for $N=2$. Solid lines gives the threshold for which $\mathcal{E}(U,N_s,N)<10^{-3}$. In figure \textbf{b} we show the threshold given by condition $\mathcal{E}(U,N_s,N)<10^{-3}$ for different number of particles and system sizes. }
 \label{fig:size_scaling}
 \end{figure}
\end{center} 
%\twocolumngrid

\section{Finite-size effects}

In order to relate the size of the many-body bound state and the periodicity of the currents we analize the dependence of the ground-state energy on the artificial gauge flux $\Omega/\Omega_0$ for various values of interaction strength $U$ and different system sizes $N_s$.

We estimate the spatial size associated to the many-body bound state by studying the exponential decay of the density-density correlations \cite{Naldesi2018}
\begin{equation}
\langle n_{j}n_{j+r} \rangle \approx \text{exp}[-r/\xi] .
\end{equation}

\begin{center}
 \begin{figure}[htb]
 \includegraphics[width=1.0 \columnwidth ]{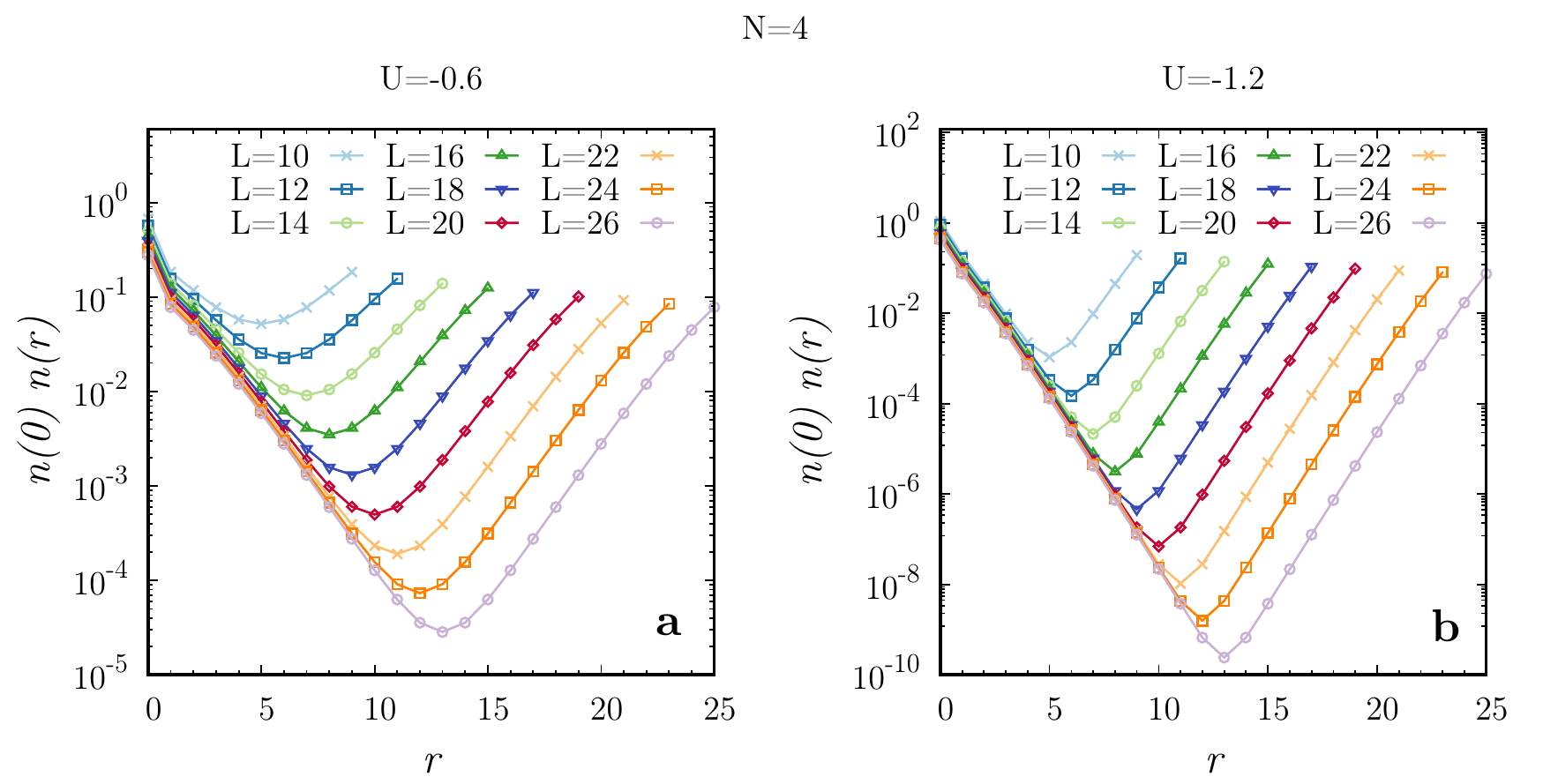}
 \caption{Density-density correlations $C_{j,j+r}$ for $N=4$ within and outside the regime where the system presents an increase of the periodicity of the current. }
 \label{fig:size_scaling_2}
 \end{figure}
\end{center}

\begin{center}
 \begin{figure}[htb]
 \includegraphics[width=1.0 \columnwidth ]{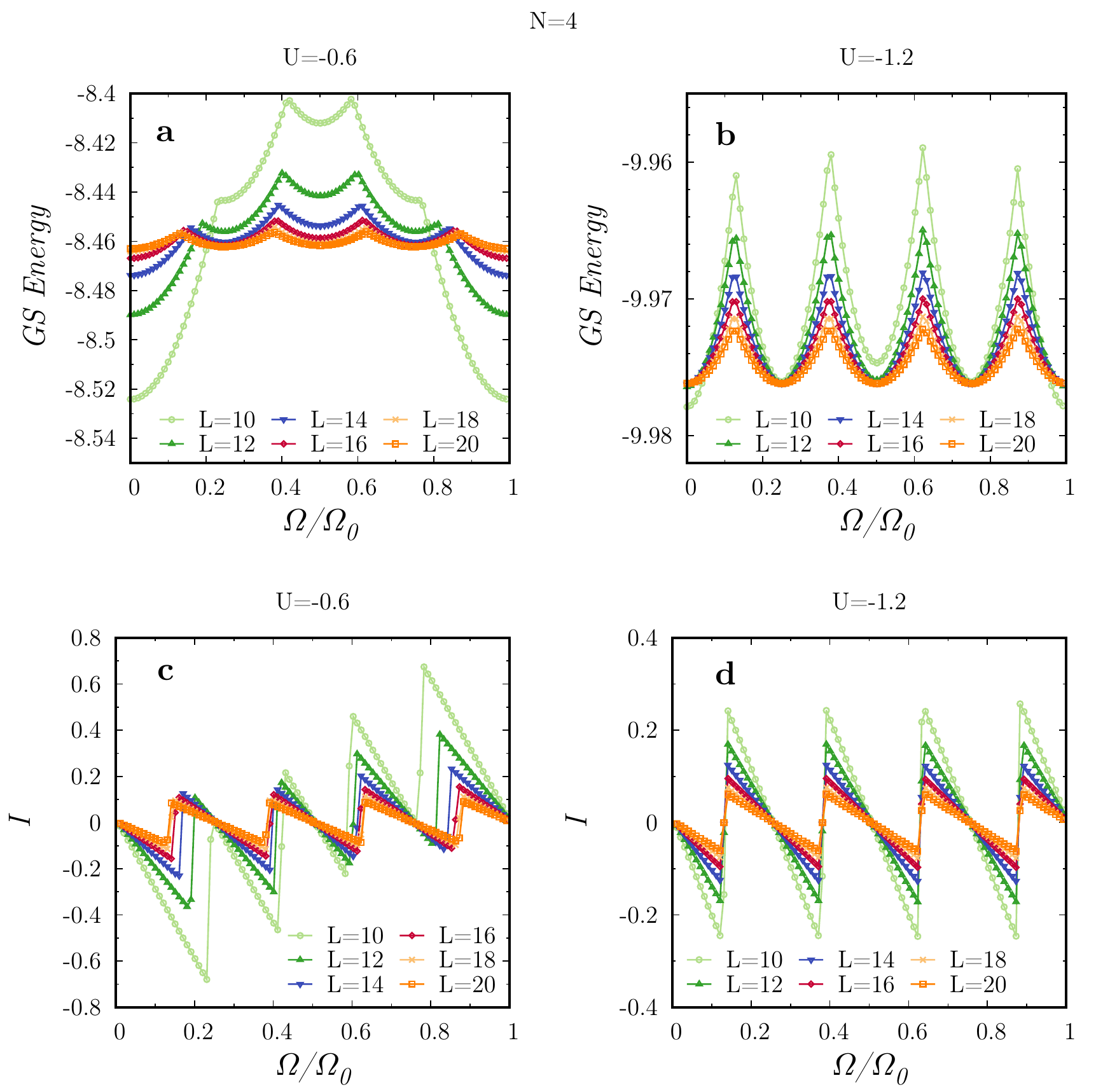}
 \caption{\textbf{a}, \textbf{b} round state energy for different interaction as a function of the flux. \textbf{c}, \textbf{d} current, calculated as $I=-\partial E_{GS}(\Omega)/\partial \Omega$, for different interaction as a function of the flux.}
 \label{fig:size_scaling_3}
 \end{figure}
\end{center}

We quantify the quality of the $1/N$ periodicity of the ground-state energy $E(\Omega)$ by calculating 
\begin{equation}
\mathcal{E}(U,N_s,N) = \frac{|E(\Omega=\Omega_0/N)-E(\Omega=0)|}{E(\Omega=0)},
\end{equation}
such that $\mathcal{E}(U,N_s,N)=0$ corresponds to a perfect $1/N$ periodicity. 
Figure (\ref{fig:size_scaling})(a) shows the density plot $\mathcal{E}(U,N_s,N)$ for a fixed number of particles $N=2$. In this figure, we show that for large $U$ and a sufficiently large system size, the periodicity of the ground-state energy is increased by a factor $N$ with respect to the noninteracting case. In Fig.~\ref{fig:size_scaling})(b) we calculated the threshold for which the minimum of the $N$-time periodicity is obtained within an error of $0.1\%$, i.e. $\mathcal{E}<10^{-3}$, for different number of particles (corresponding to the solid line in Fig.~\ref{fig:size_scaling})(a)). Finally we compare the density-density correlations $\langle n_{j}n_{k} \rangle$ for two different points in the density plot shown in (a), one within the region where the current presents $N$-time periodicity and one above the threshold. Indeed, comparing Fig.\ref{fig:size_scaling_2} and Fig.\ref{fig:size_scaling_3}, we can demonstrate that the size of the soliton, which depends on $U$ for a fixed number of particles, must be smaller than $N_s$ in order to observe the enhanced sensitivity presented in this paper.\\

\section{Numerical simulations of the quench dynamics}

We describe the many-body dynamics following the quench from $\Omega=0$ to $\Omega=\Omega_0/2$ by means of exact diagonalization. We evaluate
\begin{equation}
|\psi(t)\rangle=e^{- i \hat {\cal H}_{tot} t} |\psi(0)\rangle  
\end{equation}
where $ \hat {\cal H}_{tot}=\hat {\cal H}_{BH}+\Delta_0 \hat n_{\bar j}$ with $\hat {\cal H}_{BH}$ taken at $\Omega=\Omega_0/2$ and 
$|\psi(0)\rangle $ is the ground state  of the pre-quench Hamiltonian $\hat {\cal H}_{BH}+\Delta_0 \hat n_{\bar j}$ taken at $\Omega=0$. 
The fidelity is then obtained as ${\cal F}=|\langle \psi(t)|\psi\rangle_{NOON}|^2$
and the current is given by $I(t)=-i J\sum_j \langle \psi(t) | a_{j+1}^\dagger a_j- H.c.|\psi(t)\rangle$
The oscillation period of the current can be modified by tuning the barrier strenght.
The physical parameters we used to obtain the data shown in Fig.~\ref{fig:fisher}, panel (b) are obtained as follows: for each value of $N$ we choose $U$ in order to have the same spatial size of the many-body bound state as obtained by the study of the density-density spatial correlation function. The final choices are summarized in the Table~\ref{tab1} below.

\begin{table}[h!]
\begin{tabular}{ccccc}
$N$ 	&	\quad $U/J$  	&	\quad 	$\Delta_0/J$ 	\\ \hline
2  	&	\quad -1.06 	&	\quad	0.05			\\
3  	&	\quad -0.72 	&	\quad	0.03			\\
4  	&	\quad -0.52 	&	\quad	0.025		\\
5  	&	\quad -0.40 	&	\quad 	0.01          
\end{tabular}
\caption{Choice of parameters for the study of the quech dynamics. At varying particle numbers, we have chosen the interaction stregth and the barrier strength in such a way that the many-body bound state has the same size. All the calculations are performed with $N_s=20$.}
\label{tab1}
\end{table}

\end{document}